\documentclass[amsmath,superscriptaddress,showpacs,aps,prl,twocolumn]{revtex4-1}
\usepackage[colorlinks,linkcolor=blue,anchorcolor=blue,citecolor=blue,urlcolor=blue]{hyperref}
\usepackage{amsmath}
\usepackage{amssymb}
\usepackage{graphicx}
\usepackage{color}
\usepackage{bm}

\begin{document}
\title{Fractionalized Spin Excitations in the Edge Ferromagnetic State of Graphene: Signature of the Ferromagnetic Luttinger Liquid}

\author{Zhao-Yang Dong}
\affiliation{Department of Applied Physics, Nanjing University of Science and Technology, Nanjing 210094, China.}
\author{Wei Wang}
\affiliation{School of Science, Nanjing University of Posts and Telecommunications, Nanjing 210023, China}
\affiliation{National Laboratory of Solid State Microstructures and Department of Physics, Nanjing University, Nanjing 210093, China}
\author{Zhao-Long Gu}
\email[]{waltergu1989@gmail.com}
\affiliation{National Laboratory of Solid State Microstructures and Department of Physics, Nanjing University, Nanjing 210093, China}
\author{Shun-Li Yu}
\affiliation{National Laboratory of Solid State Microstructures and Department of Physics, Nanjing University, Nanjing 210093, China}
\affiliation{Collaborative Innovation Center of Advanced Microstructures, Nanjing University, Nanjing 210093, China}
\author{Jian-Xin Li}
\email[]{jxli@nju.edu.cn}
\affiliation{National Laboratory of Solid State Microstructures and Department of Physics, Nanjing University, Nanjing 210093, China}
\affiliation{Collaborative Innovation Center of Advanced Microstructures, Nanjing University, Nanjing 210093, China}
\date{\today}

\begin{abstract}

\par The elementary excitations from the conventional magnetic ordered states, such as ferromagnets and antiferromagnets, are magnons. Here, we elaborate a case where the well-defined magnons are absent completely and the spin excitation spectra exhibit an entire continuum in the itinerant edge ferromagnetic state of graphene arising from the flatband edge electronic states. Based on the further studies of the entanglement entropy and finite-size analysis, we show that the continuum other than the Stoner part results from the spin-1/2 spinons deconfined from magnons. The spinon continuum in a magnetically ordered state is ascribed to a ferromagnetic Luttinger liquid in this edge ferromagnet. The investigation is carried out by using the numerical exact diagonalization method with a projection of the interacting Hamiltonian onto the flat band.

\end{abstract}

\maketitle


\par Graphene is famous for being a host of Dirac fermions, and has attracted great interest over the past decade for its unique electronic properties \cite{Novoselov2005,RevModPhys.81.109}. The band structure of graphene is characterized by a pair of touching points with opposite chiralities, which are symmetry-protected at the high-symmetry points $K$ and $K'$. These points are two-dimensional analogs of Weyl nodes, which are accompanied by an emergence of one-dimensional ($1$D) edge states from one node to the other, similar to the Fermi arcs at the surface of three-dimensional Weyl semimetals \cite{PhysRevB.83.205101,Yan2017}. For finite graphene ribbons, a flat band connecting $K$ and $K'$ points crosses one third of the 1D Brillouin zone (BZ) at a zigzag boundary and crosses the other two thirds at the opposite beard boundary, while it disappears in an armchair boundary as the $K$ and $K'$ points are projected onto the same point \cite{Fujita1996,PhysRevB.54.17954,PhysRevB.59.8271,PhysRevB.77.085423}. The presence of Dirac cones in other two-dimensional semimetals also leads to homologous flatband edge states \cite{PhysRevB.82.085118,Imura2011,Matveeva2019}. Although for the bulk states of the Dirac semimetals the electron-electron interactions could be negligible, the edge states become unstable upon Coulomb interactions and traditional perturbative treatments of the interaction fail due to the absence of the dispersion.

\par Density functional theory calculations have predicted that local magnetic moments can form on the boundary of zigzag-terminated graphene nanoislands, nanodisks, and nanoribbons \cite{Fernandez-Rossier2007,Jiang2007,PhysRevB.86.195418}. The experimental evidences of the edge magnetization are also reported \cite{RamakrishnaMatte2009,PhysRevB.81.245428,Tao2011,Magda2014,Makarova2015}. Thus considering that the itinerant ferromagnetism would arise from Coulomb repulsions in flat or nearly flat bands \cite{Tasaki_PTP1998,Tasaki_PRL1992,Mielke_PLA1993,Mielke_CMP1993}, the mechanism of the edge ferromagnetism has been studied based on the flatband edge states: the work studying effective models with onsite Hubbard interactions on a zigzag boundary argued for ferromagnetic states \cite{Fujita1996,Fernandez-Rossier2007,Jung2009,Karimi2012}; weak-coupling renormalization group and density-matrix renormalization group calculation, and quantum Monte Carlo simulation also provided similar results \cite{PhysRevB.68.035432}. Given that there is no magnetic moments in the bulk of graphene, the local moments on the zigzag boundary confirmed by Lieb's theory \cite{PhysRevLett.62.1201} further validate that the edge ferromagnetic states of graphene can be interpreted by the flatband ferromagnetism \cite{Fernandez-Rossier2007}.

\par The edge ferromagnetism of graphene, as a quasi-$1$D system circumambulating the Lieb-Mattis theorem \cite{PhysRev.125.164}, is expected to be distinct from the conventional ferromagnetism. However, previous researches primarily focused on the origin and spatial distribution of the moments of the edge ferromagnetic order other than the elementary excitations that would reveal the fundamental difference.
Additionally, considering that the flatband edge states can be driven by the spin-orbital coupling into the $1$D edge states of topological insulators, where the interactions result in a Luttinger liquid \cite{PhysRevB.73.205408,PhysRevLett.96.106401,Stuhler2019,Ciftja2012}, whether the excitations in the edge ferromagnetism behave as a Luttinger liquid is an intriguing problem. Therefore, the excitation spectra deserve a theoretical study to understand the nature of the edge ferromagnetism. On the other hand, the edge magnetism of graphene offers unique opportunities for future technological applications such as spintronics \cite{Son2006,PhysRevLett.100.177207}, thus the study of its edge excitations is also significant to the spin-transport investigations.

\par In this letter, we study the spin dynamics based on the edge ferromagnetic ground state. With a projection of the interacting Hubbard model onto the flatband edge states, the dynamic spectra are calculated by use of the numerical exact diagonalization method.
Remarkably, the spin excitation spectra exhibit an entire continuum and the usual well-defined magnonic excitations are absent with only a broad dome-shaped reminiscence left. Based on the further studies of the entanglement entropy (EE) between spin-up and spin-down subspaces, we show that the reminiscence is composed of fractionalized spin-1/2 excitations. Furthermore, by comparing the spectra with those obtained on graphene ribbons with finite widths, we suggest these fractionalized excitations to be spinons deconfined from the magnons strongly coupled to the Stoner excitations. We argue that the emergence of deconfined spinons in a magnetically ordered state here indicates a ferromagnetic Luttinger liquid in the graphene edge ferromagnet.

\begin{figure}
\centering
\includegraphics[width=0.48\textwidth]{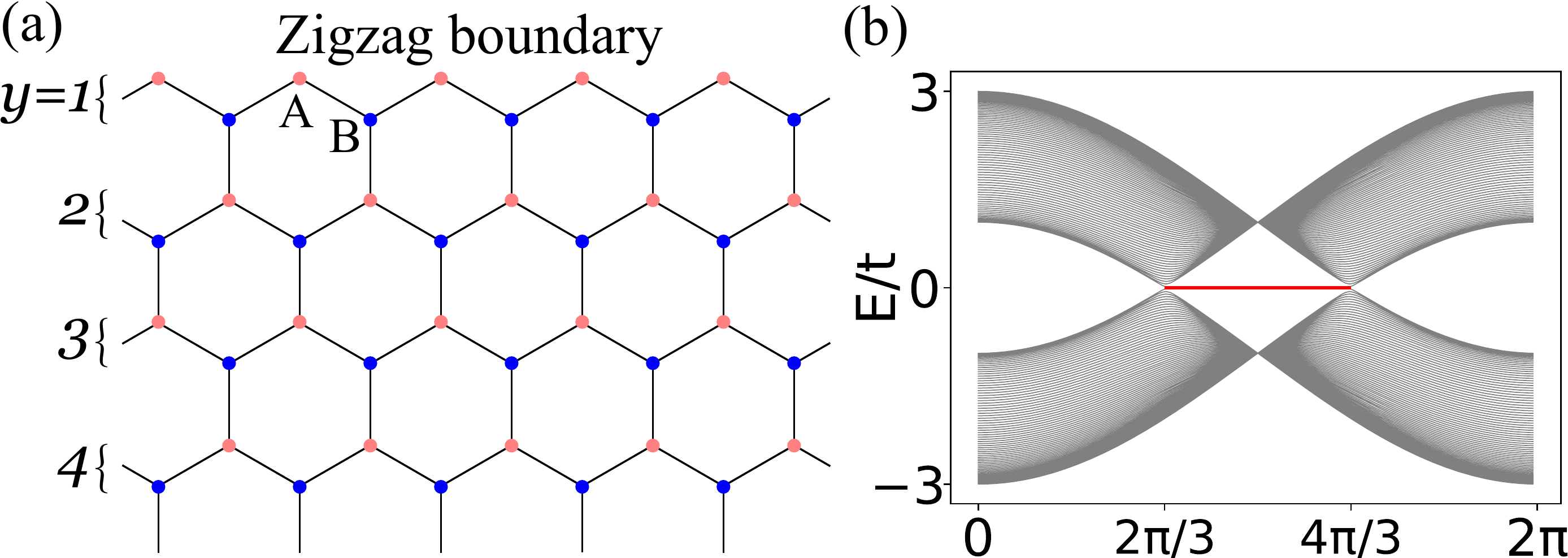}
\caption{(color online). (a) Illustration of the hexagon lattice with a zigzag boundary. The pink (blue) sites denote the A(B) sublattices. (b) Energy bands of Eq.~(\ref{H0}) with a zigzag boundary. The red line denotes the flatband edge states.}
\label{system}
\end{figure}

\par Let us begin with the tight-binding Hamiltonian for graphene, which can be simply written as $H=t\sum_{\langle ij\rangle\sigma}a^{\dagger}_{i\sigma}b_{j\sigma}+{\rm H.c.}$, where $i=(x,y)$ denotes the two-dimensional coordinates of sites and $a_{i\sigma}(b_{i\sigma})$ is a standard notation of a spinful fermionic annihilation operator for the electron on site $i$ of sublattice A(B). For simplicity, only the hopping terms over the nearest-neighbor $\langle ij\rangle$ bonds are involved.
Considering a half-infinite graphene with a zigzag boundary as illustrated in Fig.~\ref{system}(a), the periodic orientation of the Hamiltonian can be written in the momentum space,
\begin{equation}\label{H0}
H=\sum_{k\sigma}
C^\dagger_{k\sigma }
\left(
  \begin{array}{ccc}
    T_\sigma(k) & D_\sigma(k) &\cdots\\
    D^\dagger_\sigma(k) & T_\sigma(k) & \cdots \\
    \vdots & \vdots & \ddots\\
  \end{array}
\right)
C_{k\sigma'},
\end{equation}
where $C^\dagger_{k\sigma }=(a^\dagger_{1k\sigma},b^\dagger_{1k\sigma},a^\dagger_{2k\sigma},\cdots)$ is the vector of operators along the orientation of the boundary, and $T_\sigma(k), D_\sigma(k)$ are the Fourier transforms of the hopping matrix,
\begin{equation}\label{grapheneAB}
 T_\sigma(k)=\left(
     \begin{array}{cc}
       0 & t(1+{\rm e}^{-{\rm i}k}) \\
       t(1+{\rm e}^{{\rm i}k}) & 0 \\
     \end{array}
   \right),
 D_\sigma(k)=\left(
     \begin{array}{cc}
       0 & 0 \\
       t & 0 \\
     \end{array}
   \right).
\end{equation}
Solving Eq.~(\ref{H0}), as shown in Fig.~\ref{system}(b), a zero-energy flat band from $2\pi/3$ to $4\pi/3$, connecting the two Dirac cones, can be obtained, and the quasiparticles read
\begin{equation}\label{state}
  d_{k\sigma}=\sum_y \mu^\ast_{yk}a_{yk\sigma},
\end{equation}
where $\mu_{yk}=(-1-{\rm e}^{{\rm i}k})^{y-1}\sqrt{-2\cos{k}-1}$ is the probability amplitude of the electron that contributes to the flat band.
Their localizations at the boundary guarantee that they are flatband edge states. In this case, the edge ferromagnetic ground state resulting from the Hubbard interaction $H_U=U\sum_{i} n_{i\uparrow }n_{i\downarrow}$ in the flat band is expected to be $|\text{FM}\rangle\equiv\prod_{k\in\text{FBZ}}d^\dagger_{k\sigma}|\Omega\rangle$, where $|\Omega\rangle$ is the bulk ground state.
The absence of the dispersion leaves the interaction to govern the low-energy physics of the edge states, so that the physics is dominated by the degrees of freedom of the flat band and the Hubbard interaction can be projected onto it \cite{Wang2012,Neupert_PRL2011,Regnault_PRX2011,Neupert_PRB2011,Luitz2011,Neupert_PRL2012,Su_PRB2018,Su_PRB2019,Gu_arXiv2019},
\begin{eqnarray}
H_{\rm eff}=&&PH_UP \nonumber\\
=&&{U \over 2N}\sum_{y\sigma \sigma'}\sum_{kk'p} \mu^{\ast}_{yk+p}\mu^{\ast}_{yk'-p}\mu_{yk'}\mu_{yk}\nonumber\\
&&\times d^\dagger_{k+p\sigma}d^\dagger_{k'-p\sigma'}d_{k'\sigma'}d_{k\sigma},\label{Heff}
\end{eqnarray}
where $P$ is the projector onto the flat band.
Thus, a basis of spin-flip excitations with a center-of-mass momentum $q$ can be written as $|k_i\rangle_q=d^\dagger_{k_i-q\downarrow}d_{k_i\uparrow}|\text{FM}\rangle$. The matrix element of Eq.~(\ref{Heff}) on this set of bases is
\begin{equation}\label{PHP}
_q\langle k_j|H_{\rm eff}|k_i\rangle_q=M_{i}^s(q)\delta_{k_j,k_i}-M_{ji}^m(q)
\end{equation}
where
\begin{eqnarray}
M_{i}^s(q) &=& \frac{U}{N}\sum_{y}\sum_{p}\left|\mu_{yp}\right|^2\left|\mu_{yk_{i}-q}\right|^2, \label{M2}\\
M_{ji}^m(q) &=& \frac{U}{N}\sum_{y}U
\mu^{\ast}_{yk_i-q}\mu_{yk_{i}}\mu_{yk_j-q}\mu^{\ast}_{yk_{j}}. \label{M3}
\end{eqnarray}
The summation over $y$ could be limited to the sites on the boundary, where the edge states are localized on, due to the exponential decay of the amplitude $\mu_{yk}$. 
A remarkable consequence of the projection onto the flatband edge states is that the dimension of the Hilbert space of spin-$1$ excitations scales linearly with respect to the system size. This enables us to access a much larger system.

\par With Eq.~(\ref{PHP}), we can study the spin excitations over the edge ferromagnetic ground state in the projected space with the system size $N=1024$. The edge spin correlation function is
\begin{equation}\label{correlation}
  S^{+-}(q,\omega)=\frac{1}{2\pi}\int\sum_{y,y'}\langle \tilde{S}^+_y(-q,0) \tilde{S}^-_{y'}(q,t)\rangle {\rm e}^{{\rm i}\omega t}dt
\end{equation}
where  $\tilde{S}^-_y(q)=\sum_k\mu^\ast_{yk}\mu_{yk-q}d^\dagger_{k\downarrow} d_{k-q\uparrow}$ is the projected spin operator.
Then the spectral function of Eq.~(\ref{correlation}) can be obtained by
\begin{equation}\label{spectral functions}
  A^{+-}(q,\omega)=-\frac{1}{\pi}{\rm Im}[\sum_i\frac{|\langle P^+
  (q)|\Psi_i(q)\rangle|^2}{\omega-E_i(q)+{\rm i}\eta}]
\end{equation}
where $\langle P^+(q)|=\langle\text{FM}|\sum_y\tilde{S}^+_y(q)$, and $E_i(q)$ and $\Psi_i(q)$ are the eigenvalues and eigenstates of Eq.~(\ref{PHP}). $\eta$ is taken to be $0.002$ in the calculations.

\begin{figure}
\centering
\includegraphics[width=0.42\textwidth]{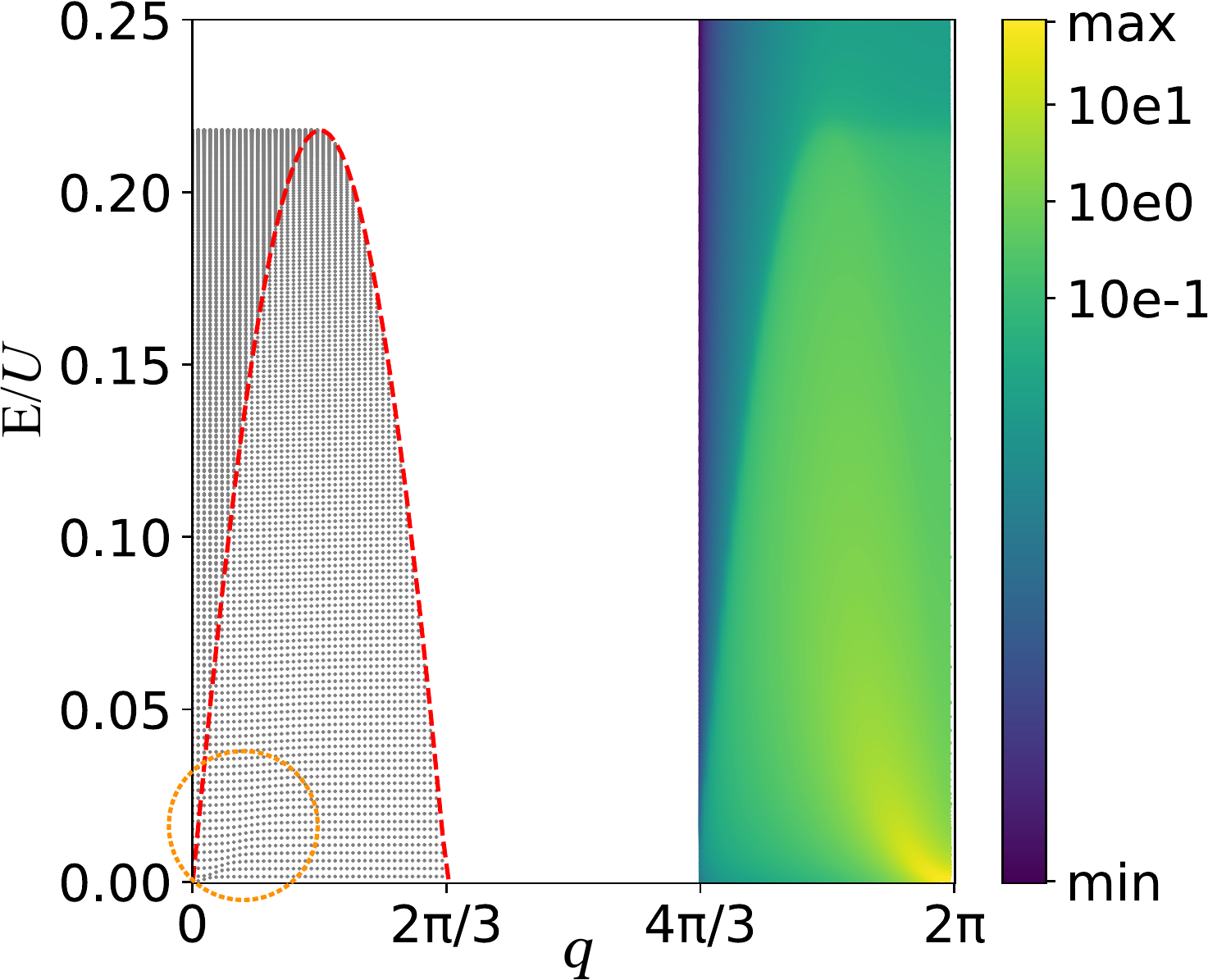}
\caption{(color online). Spin excitation energy spectra ($0\leq q\leq2\pi/3$) and spectral functions [see Eq.(9)] ($4\pi/3\leq q\leq2\pi$) over the edge ferromagnetic ground state at the zigzag boundary of a half-infinite graphene. The orange circle marks the distortions around the Goldstone mode.}
\label{spin}
\end{figure}

\par The results are presented in Fig.~\ref{spin}. Because the $2\pi/3\leq k\leq4\pi/3$ edge states only support $|q|\leq2\pi/3$ spin excitations, the spectra are distributed through $0\leq q\leq2\pi/3$ and $4\pi/3\leq q\leq2\pi$ symmetrically. In Fig.~\ref{spin}, the energy spectra are shown in $0\leq q\leq2\pi/3$ while the spectral functions are shown in $4\pi/3\leq q\leq2\pi$, respectively. As is well known \cite{Kusakabe_PRL1994,Su_PRB2018,Su_PRB2019}, the spin excitations over a flatband ferromagnet consist of two parts, the high-energy individual modes termed as the Stoner continuum and the low-energy collective modes known as the magnons. A remarkable feature of the spin excitation spectra shown in Fig.~\ref{spin} is that it exhibits an entire continuum all over the whole energy ranges. No well-defined magnonic excitations can be observed. The high-energy spectra are a dispersionless continuum, which coincide with the usual Stoner modes excited from flatband ferromagnets \cite{Su_PRB2018,Su_PRB2019}. The low-energy spectra possess more sophisticated features. In particular, as marked by the orange circle in Fig.~\ref{spin}, there exists a distortion of the dispersions in the energy spectra starting from $q=0$. This corresponds to a rather high intensity in the spectral functions around $q=2\pi$ as shown in the right part of Fig.~\ref{spin}. These ``bright'' modes extend parabolically from zero energy at $q=0$, and are merged into a dome-shaped continuum with an upper boundary shown as the red dashed line in Fig.~\ref{spin}. The boundary is related to the exact particle-hole individual modes $d_{2\pi/3-q\downarrow}^\dagger d_{2\pi/3\uparrow}|\text{FM}\rangle$, which are the eigenstates of Eq.~(\ref{PHP}). Beyond this boundary, the dispersions remain flat, thus the spectra there can be identified as the Stoner continuum. The brightest mode at $q=0$ with zero energy is the Goldstone mode resulting from the spontaneous spin SU(2) symmetry breaking. Then an issue arises as what the ``bright'' modes around $q=0(2\pi)$ are, and what their relationship to the ``disappeared'' well-defined magnons is.

\par A natural conjecture is that these ``bright'' modes are over-damped magnons. After all, they seem to preserve much of the features of the usual ferromagnetic magnons: the noticeable intensity in the spectral function and the parabolic ``dispersion relation'' around the Goldstone mode. However, we show in the following EE analysis that this possibility can be exempt. Moreover, it is found that these ``bright'' modes are in fact fractionalized spin-1/2 excitations. To see this, we note that the system can be divided into spin-up and spin-down space. That is, the $i$th eigenstate of Eq.~(\ref{PHP}) can be written as
\begin{equation}\label{densityM}
  |\Psi_i(q)\rangle=\sum_k\psi_i(k) |\downarrow_{k-q}\rangle\otimes|\Uparrow_k\rangle,
\end{equation}
where $|\downarrow_{k-q}\rangle\otimes|\Uparrow_k\rangle=|k\rangle_q$, which is represented by the direct product of a particle in spin-down space and a hole in spin-up space. Meanwhile, Eq.~(\ref{densityM}) is also the Schmidt decomposition of the state. Then, with respect to this bipartition, the EE of $|\Psi_i(q)\rangle$ now reads,
\begin{equation}\label{entropy}
  S_i(q)=-\sum_k|\psi_i(k)|^2\ln{|\psi_i(k)|^2}.
\end{equation}
When the interactions between electrons are absent, a spin excitation is just a particle-hole pair, which is a product state with respect to the above bipartition, leading to zero EE between the spin-up and spin-down subspaces. When the interactions set in, particle-hole pairs are coupled to each other and new eigen modes emerge. In general, these new modes bear more entanglement. One case is that the particle-hole pairs are confined to form bound states with the spin-up and spin-down subspaces indivisible to each other. This corresponds to the magnonic modes, whose EEs are logarithmically divergent in the thermodynamic limit. Another case is that the new modes can be decomposed into two relatively free parts with one part taking up the spin-up degrees of freedom and the other part taking up the spin-down ones. In this case the EE converges to a constant. This corresponds to fractionalized spin-1/2 excitations or individual spin excitations.


\begin{figure}
\centering
\includegraphics[width=0.4\textwidth]{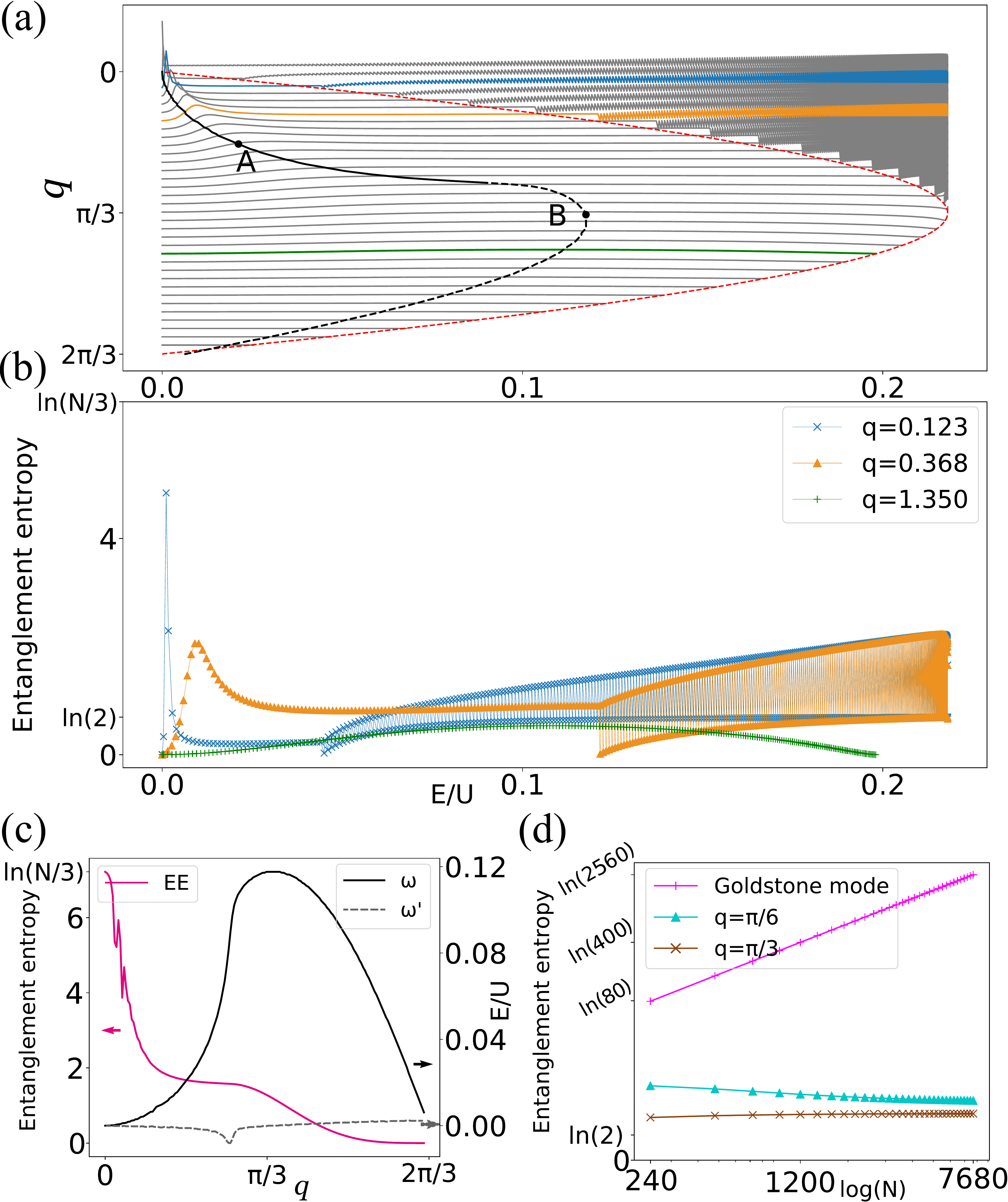}
\caption{(color online). (a) Spectra of the EEs of the spin excitations (system size $N=2048$). The black line denotes the positions of the maxima of the EEs at different momenta and its dashed segment implies the disappearance of the peaks. (b) EE at $q=0.123$ (blue), $q=0.368$ (orange) and $q=1.350$ (green), which are also denoted by the same color in (a). (c) Maxima of the EEs at different momenta (magenta), dispersion of the maxima (black), and the derivative of this dispersion (grey) (System size $N=4096$). (d) Scaling behaviors of the EEs of the Goldstone mode (pink), the excitations at $q=\pi/6$ [cyan, point A in (a)] and $q=\pi/3$ [brown, point B in (a)]. } \label{EE}
\end{figure}

\par The EEs of the edge spin excitations are shown in Fig.~\ref{EE}(a). To see the details more clearly, we specialize three EE results denoted by the colored lines (blue, orange and green) in Fig.~\ref{EE}(a) and replot them in Fig.~\ref{EE}(b). It can be found that the EEs are also separated by the same red dashed line as that in Fig.~\ref{spin}. Beyond this line, they oscillate around $\ln{2}$, which is the feature of the Stoner excitations \cite{StonerEE}. On the other hand, the ``bright'' modes correspond to the peaks observed in the EEs, which clearly bear more entanglement than the Stoner excitations. Thus, in Fig.~\ref{EE}(a), we mark the positions in the $q$-$E$ space of the maxima of the EEs with different center-of-mass momenta by the black line, which can be viewed as some kind of ``dispersion relation'' of the ``bright'' modes. In Fig.~\ref{EE}(c), we replot this dispersion as the black solid line with its derivative to $q$ as the grey dashed line. In this subfigure, the EEs of the eigen modes along this dispersion line is also plotted as the magenta line. It can be seen that this dispersion grows parabolically from $q=0$ and falls to zero near $q=2\pi/3$. In between, there exist a nondifferentiable point at $q\approx0.313$ as indicated by the peak of its derivative. As is shown by the magenta line in Fig.~\ref{EE}(c) and can be seen from Figs.~\ref{EE}(a)-(b), the peaks of the EEs fall rapidly with the approach to this point, and disappear after it. This is consistent with the mergence of the ``bright'' modes to a dome-shaped continuum as described above. The disappearance of the peaks is also implied by the dashed segment of the black line in Fig.~\ref{EE}(a). To detect whether the ``bright'' modes are magnonic ones, we choose two representative points as marked by $A$ and $B$ in Fig.~\ref{EE}(a). The scaling behaviors of the EEs with the increase of the system size $N$ (number of points in the 1D BZ) of these two points as well as the Goldstone mode are shown in Fig.~\ref{EE}(d). Apparently, for the Goldstone mode, the EE is logarithmically divergent, meaning that the Goldstone mode is indeed a bound state of particle-hole pairs. Nevertheless, for all the other ``bright'' modes, the EE converges to a constant. That is, these ``bright'' modes are not magnonic ones at all. On the contrary, they turn out to be fractionalized spin-1/2 excitations.

\begin{figure}
\centering
\includegraphics[width=0.47\textwidth]{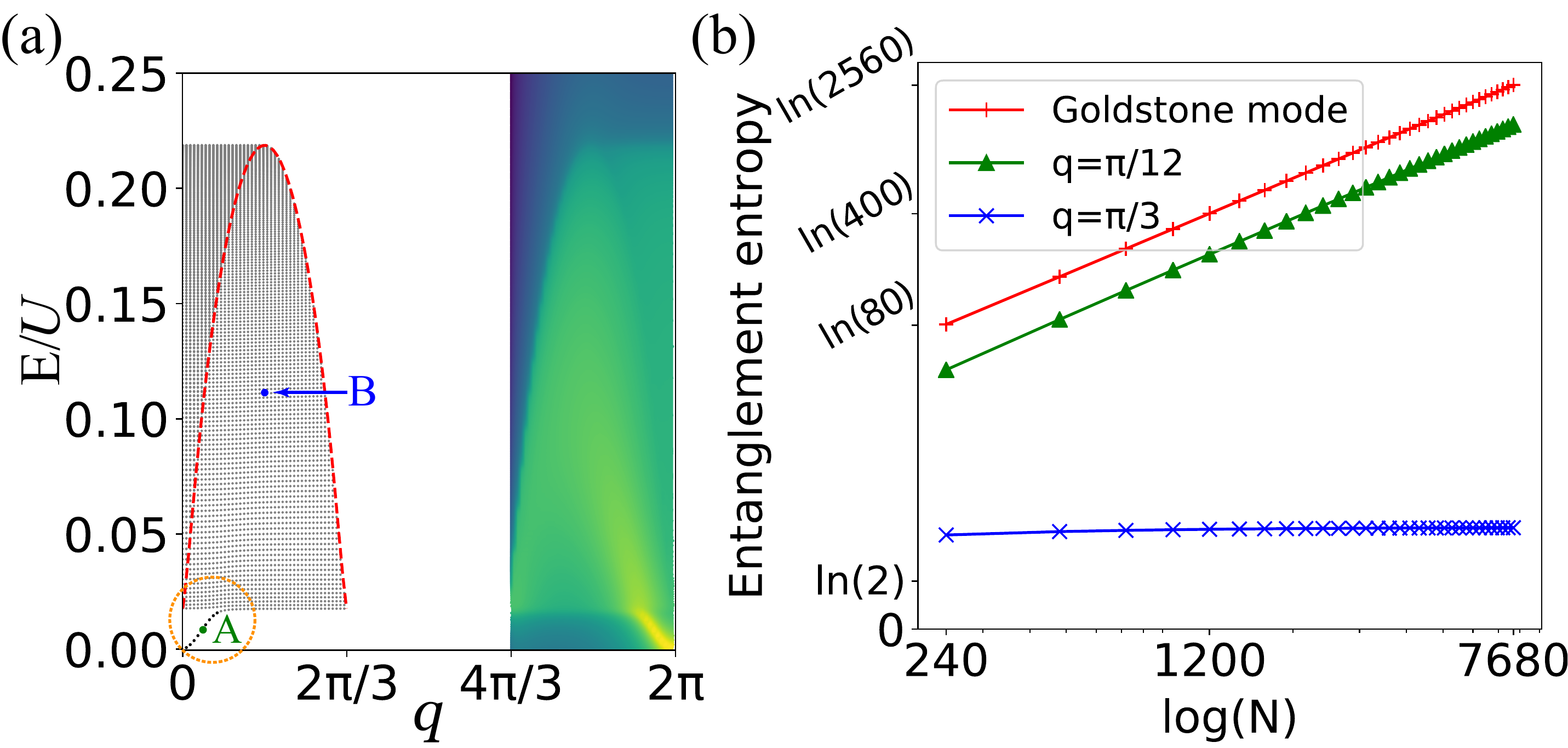}
\caption{(color online). (a) Spectra of spin excitations on the zigzag boundary of a graphene ribbon with a finite width $W=20$. Orange circle marks the well-defined magnons in the gap of the Stoner continuum. (b) Scaling of the EEs of the Goldstone mode (red), the magnonic mode at $q=\pi/12$ [green, point A in (a)], and the fractionalized mode at $q=\pi/3$ [blue, point B in (a)].}
\label{finitesize}
\end{figure}

\par To clarify the relationship between the ``bright'' modes and the ``disappeared'' magnons, in Fig.~\ref{finitesize}, we present the results obtained at the zigzag boundary of a graphene ribbon with a finite width $W=20$ for comparison. Due to the finite-size effect, as can be seen in Fig.~\ref{finitesize}(a), the Stoner continuum is gapped. Within this gap, as marked by the orange circle, well-defined magnons can be observed. Entering the Stoner continuum, the magnons are destroyed by their strong couplings to the Stoner excitations. Correspondingly, the sharp signals of the magnons become diffused and are finally merged to a dome-shaped continuum in the spectral function. In Fig.~\ref{finitesize}(b), we plot the scaling behaviors of the EEs of two eigen modes marked by $A$ and $B$ in Fig.~\ref{finitesize}(a), as the representatives of the well-defined magnons and the ``bright'' modes, respectively. The scaling behavior of the Goldstone mode is also plotted. Apparently, for both the Goldstone mode and the magnonic mode, their EEs are logarithmically divergent with the increase of the system size, while for the ``bright'' mode, its EE converges to a constant, just as expected. As the width of the ribbon grows, the gap of the Stoner continuum gets smaller and down to zero in the thermodynamic limit. Then all the well-defined magnons except the Goldstone mode will be spoiled as has been shown in Figs.~\ref{spin}-\ref{EE}. This continuous evolution indicates that the ``bright'' modes result from the fractionalization of the ``disappeared'' magnons driven by the strong couplings to the Stoner excitations, and thus are deconfined spin-1/2 spinons. Spinon continuum is a hallmark of Luttinger liquid in 1D interacting systems. Besides, a homologous deconfinement of magnons in the 1D weak itinerant ferromagnetism has been ascribed to ferromagnetic Luttinger liquids \cite{Luitz2011,Bartosch2003}. Therefore, we suggest that the ferromagnetic Luttinger liquid is realized in the edge ferromagnetism of graphene.


\par In summary, we have studied the spin excitations in the edge ferromagnetism of graphene with the numerical exact diagonalization method with a projection onto the edge flat band. Remarkably, the spin excitation spectra exhibit an entire continuum and the usual well-defined magnonic excitations are absent with only a broad dome-shaped reminiscence left. Based on the further studies of the entanglement entropy and finite-size analysis, we show that the reminiscence is composed of deconfined spin-1/2 spinons resulting from the fractionalization of the disappeared magnons. The spinon continuum in a magnetically ordered state is ascribed to a ferromagnetic Luttinger liquid in this edge ferromagnet.

\begin{acknowledgments}
\par This work was supported by the National Natural Science Foundation of China (Grants No. 11904170, No. 11674158, No. 11774152), the Natural Science Foundation of Jiangsu Province, China (Grant No. BK20190436), and National Key Projects for Research and Development of China (Grant No. 2016YFA0300401).
\end{acknowledgments}

\bibliography{reference}

\begin{thebibliography}{48}
\bibitem [{Sto()}]{StonerEE}%
  \BibitemOpen
  \href@noop {} {\bibinfo  {journal} {For a $\Delta S_z=1$ spin excitation over
  the fully polarized ferromagnetic state with $N$ electrons, the total spin
  should be $N/2-1$ except the Goldstone mode. Thus for a simple individual
  spin excitation, a pair of spin-up and spin-down particles should form a
  singlet, whose EE is $\ln{2}$}\ }\BibitemShut {NoStop}%
\end{thebibliography}

\widetext
\newpage
\appendix
\section{Supplemental Material}

\setcounter{equation}{0}
\setcounter{figure}{0}
\setcounter{table}{0}
\setcounter{section}{0}
\renewcommand{\theequation}{S\arabic{equation}}
\renewcommand{\thesection}{S\arabic{section}}
\renewcommand{\thetable}{S\arabic{table}}
\renewcommand{\thefigure}{S\arabic{figure}}

\section{The localized real-space electronic wave function}

\par
\begin{figure}
\centering
\includegraphics[width=0.7\textwidth]{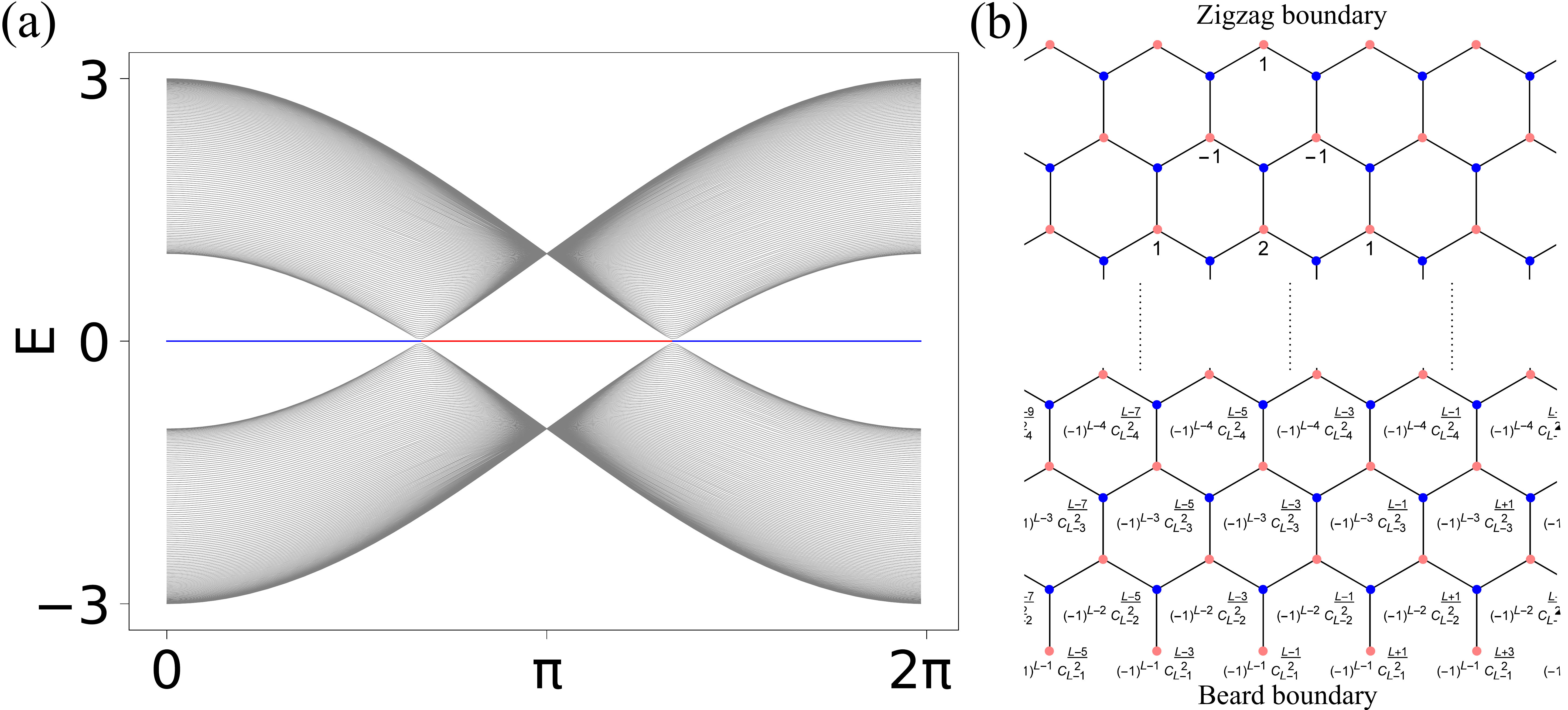}
\caption{(color online). (a) Energy bands of the graphene ribbon with the zigzag and beard boundaries. The red line denotes the states localized at the zigzag boundary and the blue line denotes the states localized at the beard boundary. (b) A graphene ribbon with the zigzag and beard boundaries. The distribution of the single-particle real-space wave function is also illustrated in the lattice. }
\label{band}
\end{figure}

 It is known that, for graphene, the flatband edge states are not only reside on the zigzag boundary but also the beard boundary, and the two parts of the edge states get together to be a whole flat band through the 1D Brillouin zone. As shown in Fig.~\ref{band}(a), the flat band connecting $K$ and $K'$ points crosses one third of the 1D Brillouin zone at a zigzag boundary denoted as the red ones, and crosses the other two thirds at the opposite beard boundary denoted as the blue.

\par The origin of flatband ferromagnetism is based on specific localized real-space electronic wave functions that overlap with each other \cite{Tasaki_PTP1998,Wang2012}. Then by an energy penalty on the overlap of two wave functions on the same site, the Hubbard repulsion lifts the anti-symmetric spin state and makes the symmetric spin state the ground state, i.e. the ferromagnetic ground state. Thus we verify the existence of localized single-particle real-space electronic wave functions to support the itinerant ferromagnetic ground state.

\par By a Fourier transform $\Psi(r)=1/\sqrt{N}\sum_k{\rm e}^{{\rm i}kr}\Psi(k)$, the single-particle real-space electronic wave functions of the edge states can be obtained from the edge states on the zigzag and the beard boundaries.
The distribution of the real-space wave function is shown in Fig.~\ref{band}(b). The distribution is a Pascal's triangle, whose base is along the beard boundary. The overlap between two wave functions, by which the ferromagnetic ground state is guaranteed, would be zero when the two wave functions are separated at a certain distant for a finite-width ribbon, so there is a transition to paramagnetic phase at a critical doping \cite{Chen2017} due to the statistical mechanics \cite{Maksymenko2012}. If the ribbon is reduced to be a chain, the single-particle real-space electronic wave function would be similar to that of the Tasaki model, and any finite doping would induce a ferro-para phase transition in the thermodynamic limit.

\begin{figure}
\centering
\includegraphics[width=0.7\textwidth]{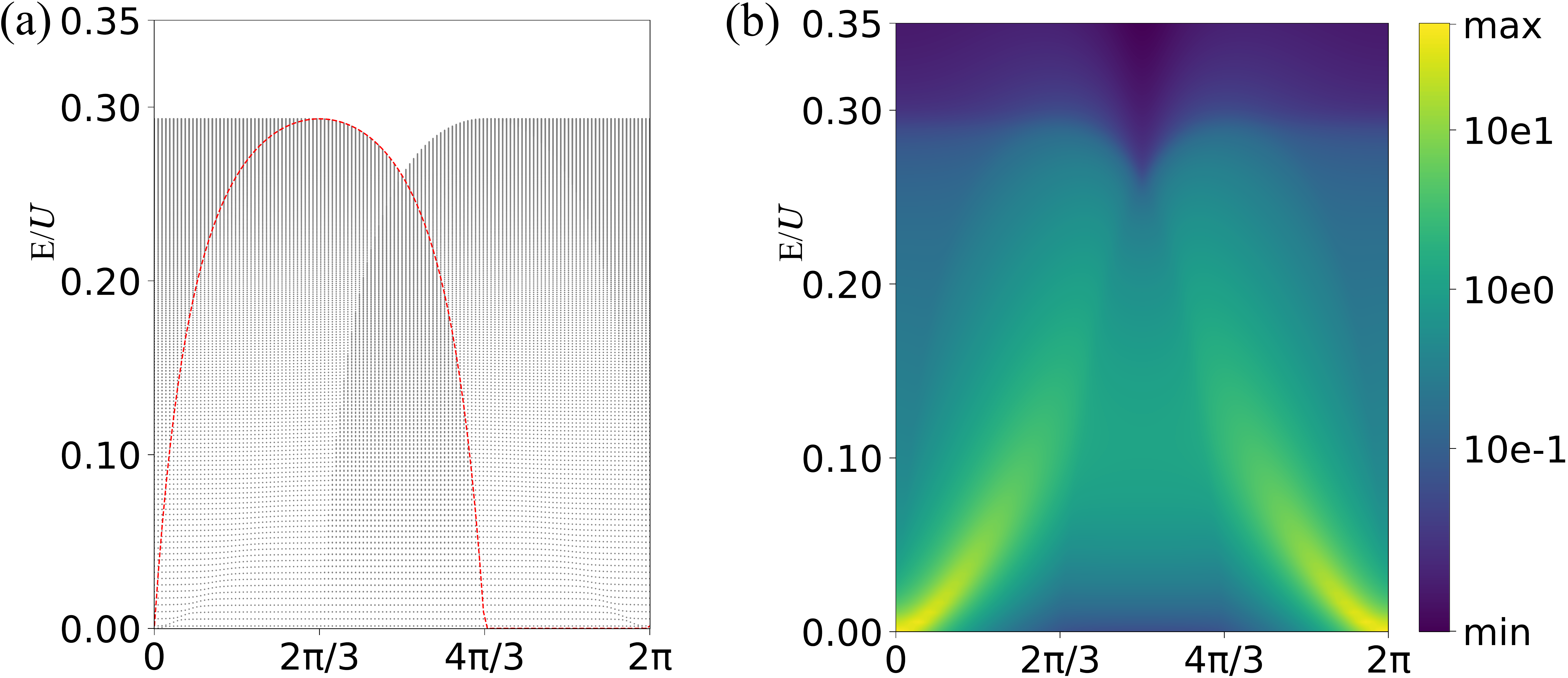}
\caption{(color online). (a) The energy spectra, and (b) the spectral functions of the spin excitations over ferromagnetic edge ground state at the beard boundary.}
\label{beard}
\end{figure}

\section{Beard boundary}

\par As mentioned in the above section, both of the flatband edge states on the zigzag and beard boundaried support the edge ferromagnetic ground states. We have illustrated the spectra of spin excitations over the zigzag boundary in the main context. Here we exhibit the other situation of beard boundary. For the half-infinite graphene with the beard boundary, the edge states are distributed in $0\leq k\leq2\pi/3$ and $4\pi/3\leq k\leq2\pi$ as shown in Fig.~\ref{band}(a). Thus the spin excitation spectra are distributed through $0\leq q\leq4\pi/3$ and $2\pi/3\leq q\leq2\pi$ symmetrically, as shown in Fig.~\ref{beard}(a) and (b). Similar to that of zigzag boundary, the magnons are also deconfined to be a spinon continuum. Besides, ``bright'' modes around the Goldstone mode imply the reminiscence of the magnons.

\section{The finite-width ribbon}

\par The bulk energy bands of a graphene ribbon will open a gap due to the finite size effect. Meanwhile two edge states at the Dirac points will emerge due to the gap at the Dirac points, and these states distributes uniformly both in the bulk and the boundaries. In this case, the effective interactions in the edge states would not be zero while approaching the Dirac points as that of the half infinite lattices. Therefore, the Stoner continuum would also open a gap due to the finite size effect. We have shown that the magnons are corrupted down due to the strong coupling to the Stoner continuum in the main context, and for the finite-width ribbon, the magnons are expected to be stable in the Stoner continuum gap.

\begin{figure}
\centering
\includegraphics[width=0.8\textwidth]{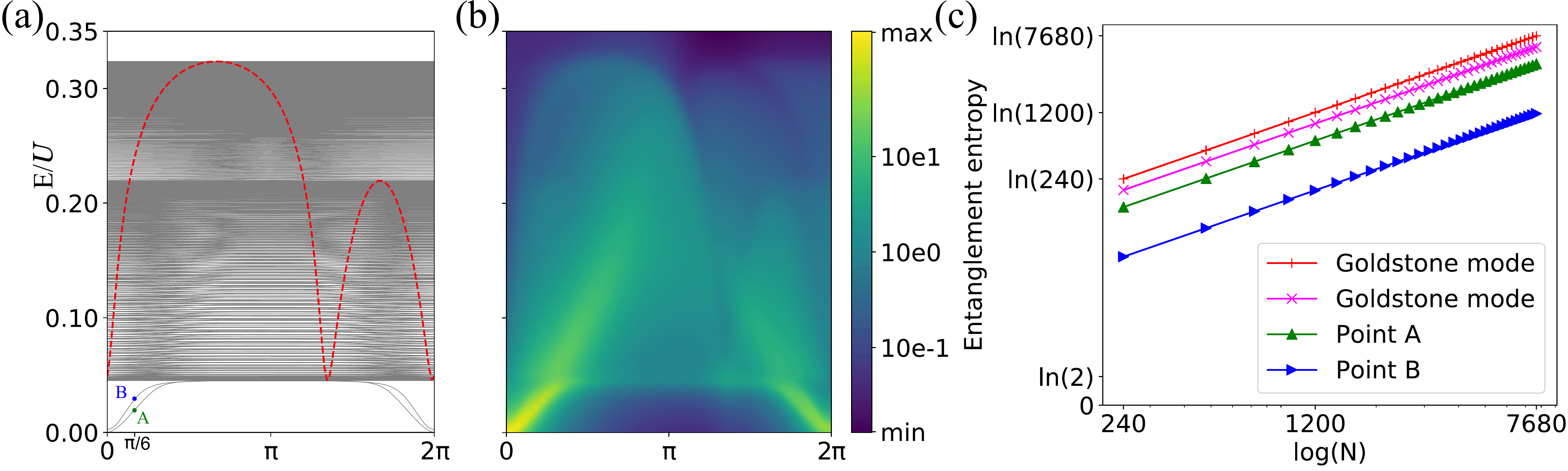}
\caption{(color online). (a) The energy spectra, and (b) the spectral functions of the spin excitations for a $W=20$-width ribbon with the zigzag and beard boundaries. (c) Scaling of the entanglement entropies of the Goldstone modes (red and magenta) and the two excitations (green and blue) at $q=\pi/6$ denoted as A and B in (a).}
\label{finite}
\end{figure}

\par In the main context, we only project onto the edge states on the zigzag boundary of a ribbon with finite width to specify the influence of the finite size effect on the spin excitations of the zigzag boundary. Here we consider the complete edge states on both zigzag and beard boundary for a ribbon. As shown in Fig.~\ref{finite}(a) and (b), there are one branch of magnons for each boundary in the finite-size gap, and both of them are spoiled to be a continuum beyond the gap. Therefore it is believed that the low-energy continuum are mainly composed by the spinons deconfined from the magnons.

\par The spectra are a simple superposition of the excitations on the two boundaries. The finite size gap is large than that of Fig.~4 in the main context due to the interactions between the different boundaries, which are ignored in Fig.~4. Nevertheless, the results shown in Fig.~4 is credible and the corresponding discussions are effective.

\par Without the coupling to the Stoner continuum, the magnons remain well-defined in the gap. To identify the magnons clearly, we calculate the entanglement entropy of the two lowest excitations at $q=\pi/6$. As shown in Fig.~\ref{finite}(c), there is no doubt that the scaling of the entanglement entropy of two excitations are so logarithmic divergences that they are well-defined magnons.

\par The summation of $y$ in Eq.~(6) and (7) is from $1$ to $W$ for the finite-width ribbon in the calculation.  It is due to the sublattice structure that the intensity at $q=0$ is stronger than that at $q=2\pi$. The summation is limited to $y=1$ for the half-infinite lattices in the main context. Compared with the spectra of the finite-width ribbon, the spectra of the edge ferromagnetism calculated in the main context are the same quanlitively except the finite-size gap, and we believe that the limitation affects the spectra little.


\end{document}